# Analysis of the rate of force development reveals high neuromuscular fatigability in elderly patients with chronic kidney disease

Antoine Chatrenet[1,2]* , Giorgina Piccoli[2], Jean Michel Audebrand[3], Massimo Torreggiani[2], Julien Barbieux[4], Charly Vaillant[3], Baptiste Morel[5], Sylvain Durand[1] & Bruno Beaune[1]

[1]Le Mans Université, Movement – Interactions – Performance, MIP, UR4334, Le Mans, France; [2]Department of Nephrology, Centre Hospitalier Le Mans, Le Mans, France; [3]Department of Endocrinology, Centre Hospitalier Le Mans, Le Mans, France; [4]Department of Digestive Surgery, Centre Hospitalier Le Mans, Le Mans, France; [5]Inter-University Laboratory of Human Movement Biology (EA 7424), Université Savoie Mont Blanc, Chambéry, France

## Abstract

**Background**  Chronic kidney disease (CKD) induces muscle wasting and a reduction in the maximum voluntary force (MVF). Little is known about the neuromuscular fatigability in CKD patients, defined as the reduction of muscle force capacities during exercise. Neuromuscular fatigability is a crucial physical parameter of the daily living. The quantification of explosive force has been shown to be a sensitive means to assess neuromuscular fatigability. Thus, our study used explosive force estimates to assess neuromuscular fatigability in elderly CKD patients.
**Methods**  Inclusion criteria for CKD patients were age ≥ 60 years old and glomerular filtration rate (GFR) < 45 mL/min/1.73 m$^2$ not on dialysis, and those for controls were GFR > 60 mL/min/1.73 m$^2$, age and diabetes matched. The fatigability protocol focused on a handgrip task coupled with surface electromyography (sEMG). Scalars were extracted from the rate of force development (RFD): absolute and normalized time periods (50, 75, 100, 150 and 200 ms, $RFD_{50}$, $RFD_{75}$, $RFD_{100}$, $RFD_{150}$ and $RFD_{200}$, respectively), peak RFD ($RFD_{peak}$ in absolute; $NRFD_{peak}$ normalized), time-to-peak RFD (t-$RFD_{peak}$) and the relative force at $RFD_{peak}$ (MVF-$RFD_{peak}$). A statistical parametric mapping approach was performed on the force, impulse and RFD–time curves. The integrated sEMG with time at 0–30, 0–50, 0–100 and 0–200 ms time intervals relative to onset of sEMG activity was extracted and groups were compared separately for each sex.
**Results**  The cohort of 159 individuals had a median age of 69 ($9_{IQR}$) years and body mass index was 27.6 ($6.2_{IQR}$) kg/m$^2$. Propensity-score-matched groups balanced CKD patients and controls by gender with 66 males and 34 females. In scalar analysis, CKD patients manifested a higher decrement than controls in the early phase of contraction, regarding the $NRFD_{peak}$ ($P = 0.009$; $\eta_p^2 = 0.034$) and $RFD_{75}$ and $RFD_{100}$ (for both $P < 0.001$; $\eta_p^2 = 0.068$ and $0.064$). The one-dimensional analysis confirmed that CKD males manifest higher and delayed neuromuscular fatigability, especially before 100 ms from onset of contraction. sEMG was lower in CKD patients than controls in the 0–100 ms (at rest: $P = 0.049$, Cohen's $d = 0.458$) and 0–200 ms (at rest: $P = 0.016$, Cohen's $d = 0.496$; during exercise: $P = 0.006$, Cohen's $d = 0.421$) time windows. Controls showed greater decrease of sEMG than CKD patients in the 0–30 ms ($P = 0.020$, Cohen's $d = 0.533$) and 0–50 ms ($P = 0.010$, Cohen's $d = 0.640$) time windows. As opposite to females, males showed almost the same differences between groups.
**Conclusions**  Our study is the first to show that CKD patients have higher fatigability than controls, which may be associated with an impaired motor-unit recruitment, highlighting a neural drive disturbance with CKD. Further studies are needed to confirm these findings.

# Introduction

Patients with chronic kidney disease (CKD) frequently have kidney disease cachexia,[1,2] resulting in a global reduction of muscle performance. In clinical routine, muscular function, as reflected by maximum voluntary force (MVF), is commonly assessed using a handgrip test, whose accuracy and reproducibility with a portable device are satisfying.[1] Its use is advised in clinical practice for CKD patients,[3] and reduced muscle function in the general population is closely associated with all-cause mortality, a risk that is exacerbated in the presence of CKD.[4]

An important aspect of human movement is the evolution of muscle performance during a fatiguing task. Neuromuscular fatigability is defined as any exercise-induced decrease in muscle performances.[5,6] In some chronic diseases, such as cancer, neuromuscular fatigability is higher than in controls.[7,8] The relationship between neuromuscular fatigability and disease-related symptoms is an important target for clinical research in order to propose specific interventions.[9,10] According to the McComas model, a decreased maximal strength does not imply an increased rate on neuromuscular fatigability.[11] To date, few studies assessed neuromuscular fatigability in elderly CKD patients.[10,12] Neuromuscular fatigability is commonly assessed by measuring MVF evolution during a standardized task[13]; however, power and speed or accuracy can be used.[9]

As explained by Buckthorpe and Roi (2017), the force production quantified using MVF does not allow us to carry out comprehensive neuromuscular testing. In daily life, several movements are performed in a few milliseconds, as can be observed during sports events or the stabilization of the body after loss of balance,[14] while the time to reach maximal strength can be longer.[15] Thus, the ability to generate force within a short interval of time (i.e., explosive contraction) has recently emerged as an important element in the assessment of muscular function in vivo.[15–17]

Rapid (explosive) force generation, defined as the rate of rise in contractile force,[18] is commonly expressed as rate of force development (RFD; see previous studies[15,16] for an extensive review of the subject). RFD relies on a range of physiological mechanisms, which themselves depend on the duration of the contraction.[19,20] In the initial phase of contraction (<100 ms), neural activation is the most important component of RFD,[18,20] while the distal strength capacities of the muscle (i.e., its cross-sectional area and architecture) are increasingly involved in the later phases, especially after 100 ms.[15,16,21] Moreover, peripheral/central components of decrease of RFD capacities during exercise may be different than aetiology of fatigability assessed during maximal sustained isometric contraction.[22] Surface electromyography (sEMG) is commonly used to assess muscle fibre recruitment during the early phase of muscle contraction.[18,19] For instance, decrease of RFD following fatiguing exercise has been related to a decrease of sEMG amplitude in the first millisecond of the contraction.[23] It is noteworthy that specific attention is required with the a priori definition of the time windows because it can induce multiple comparisons and predispose to a conservative $\alpha$ correction.[24]

Explosive force generation is sensitive to age-related changes and differs between males and females because of differences in neuromuscular functions and structures.[25] It also depends on health status. D'Emanuele et al. (2021) validated the quantification of explosive force (i.e., assessment of the RFD rather than MVF) to assess neuromuscular fatigability.[17] They revealed that explosive force assessment is more sensitive than MVF in detecting neuromuscular fatigability.

Thus, evidences of increased neuromuscular fatigability in CKD patients are lacking,[10] probably due to the low sensitivity of the test employed. To try to fill this knowledge gap, the purpose of our study was to use explosive force estimates and sEMG to assess the neuromuscular fatigability in elderly CKD patients in a clinical context, as compared with controls matched for gender, age and diabetes status. The present study sought to avoid the limitations inherent in explosive force quantification by using a statistical parametric mapping (SPM) approach.[24] In addition, sEMG analysis was chosen as a non-invasive method to support interpretation of RFD impairment in CKD patients.

# Materials and methods

## Study design

The study is prospective, single centre and observational. It is a part of the PIONEER project (PhysIOpathology of NEuromuscular function rElated to fatigue in chronic Renal disease in the elderly[26]).

## Participants

Enrolment was performed in the 'Unité d'Insuffisance Renale AVancée' at Centre Hospitalier Le Mans, in Sarthe (France), between July 2020 and August 2021.

Inclusion criteria for the CKD cohort were age over 60, estimated glomerular filtration rate (eGFR) below 45 mL/min/1.76 m$^2$ calculated by means of the Chronic Kidney Disease Epidemiology Collaboration equation for at least 3 months and stable clinical condition defined as not having been hospitalized in the previous 3 months. Inclusion criteria for controls were age of 60 or older, an available serum creatinine measurement within the previous 6 months and an eGFR above 60 mL/min/1.76 m$^2$.

Exclusion criteria were inability to give informed consent, being under guardianship, presence of neuromuscular dis-ease, known severe cognitive impairment, history of upper limb surgery and estimated life expectancy of <3 months, and those for CKD patients were an acute kidney disease or an expected start of dialysis within 3 months.

### Experimental protocol

The protocol focused on the assessment of the dominant finger flexor muscles, the muscle group most commonly studied in the clinical setting,[3] which shows high reproducibility in RFD assessment.[27] The participant sat upright with their elbow bent at 90° close to their chest. The humerus was vertical and the horizontally placed forearm rested on a height-adjustable support, positioned so the participant was able to grip the dynamometer their dominant hand. The non-dominant arm rested on the participant's leg.

### Data acquisition

The dynamometer used was the K-Force Grip (Kinvent Biomecanique SAS, Montpellier, France) with a sampling rate of 1000 Hz and accuracy of 100 g. It is fixed to a specially made support that ensures that the hand is correctly placed. The sEMG activity was recorded using Trigno® Wireless Biofeedback System (Delsys Inc., MA, USA). Sensors are composed of two pairs of silver bar contacts with 10-mm inter-electrode distance and a sampling rate of 1926 Hz. Before electrode placement, the skin was shaved and wiped clean with 70° alcohol. Four sensors were placed on the *Flexor Digitorum Superficialis* and hold with a double-sided tape and kinesiology taping while following the SENIAM recommendations and the previous anatomical location described.[26]

The participant gets direct visual feedback of the force and the sEMG by means of an interface created through LabVIEW v19.0.1 (National Instruments Corp., Austin, TX, USA). The targeted force of the submaximal contractions (40 ± 10% MVF) is represented as a coloured horizontal band in the force–time graphic.

### Familiarization and determination of the muscular parameters at rest

The protocol began with a warm-up phase that consisted of dynamic extensions of the fingers. Then, five explosive contractions defined as rapid force productions directly followed by a relaxation time[28] were performed. Each contraction lasted ~1 s and was followed by a 20-s rest. The instructions given were 'Grip the dynamometer as fast and as hard as possible and release it immediately afterwards' and the researcher gave the order to start contraction after a 3-s countdown. These explosive contractions were performed independently of the maximum voluntary contractions as recommended.[16] If a volunteer failed to reach 40% MVF or decreased in force within 200 ms from onset, the attempt was discarded. The highest RFD, calculated on the highest value of the first derivative of the force–time curve of the contraction, was selected for analysis.

After the last 20-s rest, the participant performed three maximum voluntary contractions along vigorous vocal encouragements. Contractions lasting 5 s and 2-min rest intervals between them were observed. The instructions given were 'Grip the dynamometer as hard as you can until I tell you to stop'. The researcher gave the order to start contraction after a 3-s countdown and a 5-min break was scheduled after the third attempt. The best of each participant's three attempts was considered for analysis.

### Fatigability protocol

The fatigability protocol, drawn up in accordance with Bigland-Ritchie and Woods (1984), consisted in nine submaximal contractions at 40% MVF, followed by one explosive contraction and one maximum voluntary contraction.[13] These 11 contractions were repeated 6 times. Each contraction lasted 3 s and was followed by a 2-s rest. The 40% MVF threshold was chosen because (a) it is in line with the first neuromuscular fatigability thresholds above critical force (i.e., considered as the critical threshold for neuromuscular fatigability development) and (b) neuromuscular fatigability is task specific and this threshold allows to balance central (i.e., above the neuromuscular junction) and peripheral fatigue (i.e., below the neuromuscular junction) compared with a higher threshold.[29]

### Data processing

A numeric force signal is filtered with a second-order zero-lag Butterworth lowpass filter at 40-Hz cut-off frequency with Matlab R2018a v9.4 (The MathWorks Inc., Natick, MA, USA). The onset of each contraction is determined by means of a second derivative method, described by Soda et al. (2010).[30] Force signal is considered from the onset within a 200-ms time window.

Moreover, the integral of the force–time curve ($\int force \cdot dt$), namely, impulse (N·s), is calculated at each time point in the 200-ms force continuum. Impulse incorporates the entire time history of the contraction, making identification of time-related force parameters possible.[18] The RFD–time curve is calculated through the first derivative of the force signal ($\Delta force/\Delta time$) at each overlapping time interval in the 200-ms force continuum and lowpass filtered at 50-Hz cut-off frequency.[20] Each force signal and its transformations are analysed as an absolute value and normalized by MVF.

For each absolute and relative RFD–time curve, RFD was considered at the following fixed time points: 50, 75, 100, 150 and 200 ms ($RFD_{50}$, $RFD_{75}$, $RFD_{100}$, $RFD_{150}$ and $RFD_{200}$, respectively). In addition, the following scalars are extracted (Figure 1) of the RFD–time curve[15]: the peak of the RFD–time curve expressed in absolute and relative units ($RFD_{peak}$ [N/s] and $NRFD_{peak}$ [%MVF/s], respectively), the time-to-peak RFD relative to onset of contraction (t-$RFD_{peak}$) and the percentage of MVF at $RFD_{peak}$ (MVF-$RFD_{peak}$).

The sEMG signal was filtered using a fourth-order Butterworth bandpass filter with a 10- to 500-Hz cut-off frequency. The sEMG signal-to-noise ratio was calculated during the maximal voluntary contractions to consider a single sensor for analysis. The sEMG of the explosive contractions was rectified and filtered using a zero-lag lowpass filter with 20-Hz cut-off frequency. This signal was normalized with the maximum sEMG ($EMG_{max}$) estimated on the contraction considered for the MVF, which was previously smoothed with a lowpass filter at 10-Hz cut-off frequency. As recommended,[18] the integrated sEMG with time (%$EMG_{max}$·s) was calculated within 0–30, 0–50, 0–100 and 0–200 ms time intervals relative to onset of sEMG activity. This onset was manually selected using a two-step method as previously recommended.[31] Experimental traces of CKD and control volunteers are shown in Figure 2.

## Statistical analysis

Data were tested for normality using the Shapiro–Wilk test and presented as recommended with mean and standard deviation or median and interquartile range (IQR). Normally distributed data were compared using an independent Student's t-test (e.g., CKD patients vs. controls); otherwise, the Wilcoxon rank sum test was used. Qualitative data were compared using the $\chi^2$ test or Fisher's exact test in cases of low sub-sample cohort size (i.e., <5 individuals).

Given the heterogeneity in RFD between elderly males and females,[25] analysis was performed separately for each sex. Due to the imbalance in baseline gender distribution between cases and controls, CKD patients and controls were balanced using propensity-score matching to ensure the assumption of sphericity (Figure S1) was being met. Matching was based on age and diabetes status using a Greedy algorithm with a 1:1 ratio and performed using the *MatchIt* package in RStudio (R Core Team 2021, Vienna, Austria).

The scalars, or zero-dimensional data ($RFD_{peak}$, $NRFD_{peak}$, t-$RFD_{peak}$ and MVF-$RFD_{peak}$, sEMG), were compared using SPSS v24 (IBM Corp., Armonk, NY, USA) by group (CKD patients vs. controls) and gender (female vs. male), while interactions during the fatigability period were compared using two-way repeated measures analysis of variance (ANOVA). Mauchly's test was used to verify the assumption of sphericity and, according to the epsilon statistic, the Greenhouse–Geisser or Huynh–Feldt correction was applied if required. The partial eta squared ($\eta_p^2$) with ANOVA calculation or Cohen's *d* with Student's tests were also reported to give an indication of effect size, considered as small ($\eta_p^2 \geq 0.01$, Cohen's $d \geq 0.2$), medium ($\eta_p^2 \geq 0.06$, Cohen's $d \geq 0.5$) and large ($\eta_p^2 \geq 0.14$, Cohen's $d \geq 0.8$) effects. A post hoc with Bonferroni's correction was performed. Univariate regression analysis was performed to test the relationship between scalars and eGFR.

To overcome the scalar limitation bias caused by the hypothesis definition used in this study (see Pataky et al., 2013, for additional information on technical justification and procedure[24]), we performed SPM analysis to compare one-dimensional force–time curves between groups. SPM procedures were performed using Matlab with *spm1d* package (https://spm1d.org). Normality assumption was tested using the D'Agostino–Pearson $K^2$ test and, in case of normally distributed continua, two-way repeated measures ANOVA was employed to test time (from the explosive contraction performed at rest to the last explosive contraction performed in the fatigability period) and group effect (CKD patients vs. controls). Otherwise, statistical non-parametric mapping (SnPM) with 10 000 iterations was preferred. A planned post hoc analysis was performed using Student's *t*-test with Bonferroni's correction regarding group effect (comparison of CKD patients and controls at each contraction performed in the protocol) and time effect through comparison of all the explosive contractions between the other ones in each group (e.g., in CKD patients, rest vs. the third explosive contraction). The two-sided alpha risk was set at 5%.

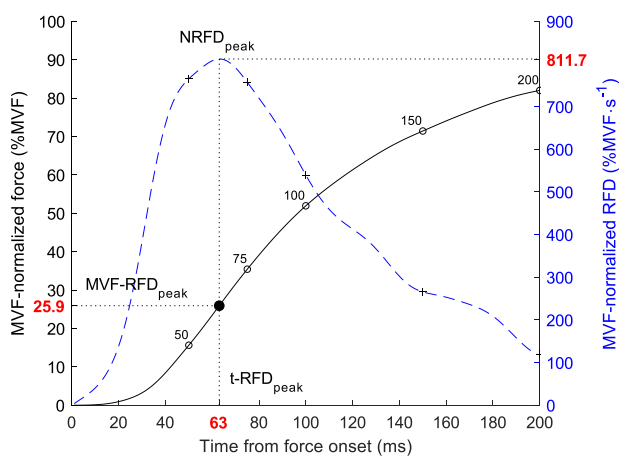

**Figure 1** Description of the scalars extracted. The black line represents the MVF-normalized force–time curve (%MVF), and the blue dashed line represents the instantaneous rate of force development (RFD) curve computed as the time derivative of the MVF-normalized force. The $NRFD_{peak}$ was determined through the peak of the MVF-normalized RFD–time curve (the peak of the absolute RFD–time curves is called $RFD_{peak}$, not represented). The percentage of the peak force at $NRFD_{peak}$ (MVF-$RFD_{peak}$) and the time from onset to $NRFD_{peak}$ (t-$RFD_{peak}$) were calculated and are symbolized by the filled black circle. The RFD considered at time periods ($RFD_{50}$, $RFD_{75}$, $RFD_{100}$, $RFD_{150}$ and $RFD_{200}$) were represented with cross symbols in the blue dashed line, with respect to the corresponding time point of the force–time curve (open black circles).

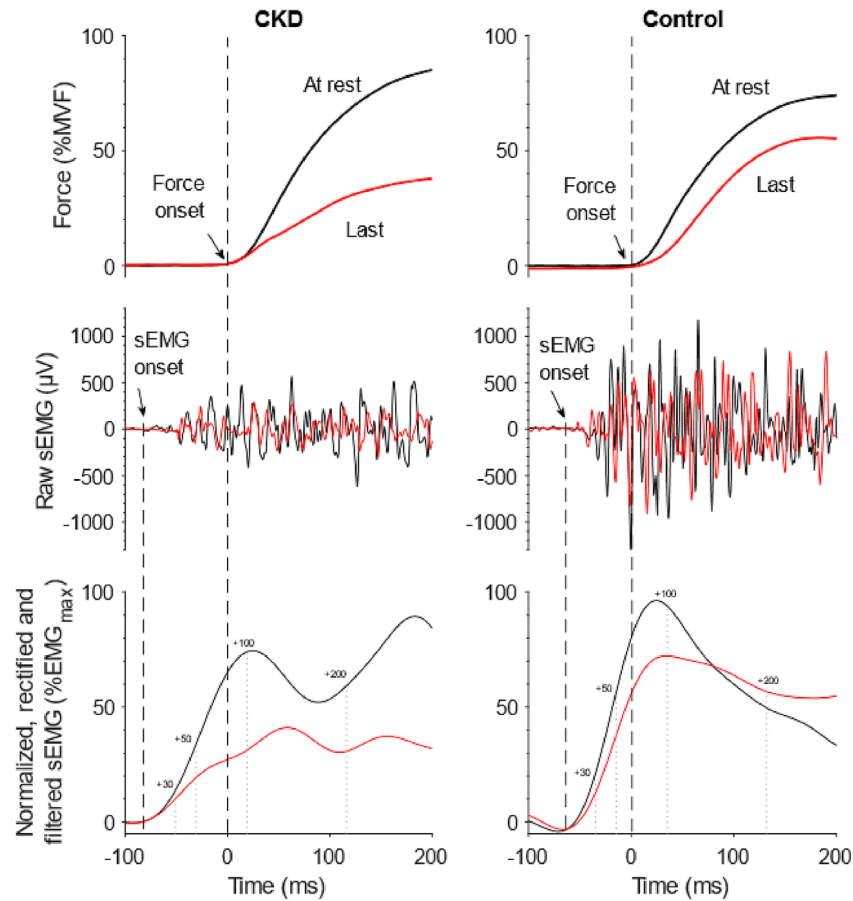

**Figure 2** Experimental traces in PRE (black) and POST exercise (red) of force and surface electromyography (sEMG) signal of representative chronic kidney disease (CKD) and control volunteers. Black dashed lines represent the detection of force and sEMG onset; grey dotted lines represent the time periods considered in sEMG integration (expressed in millisecond from sEMG onset).

## Results

### Baseline data

A total of 159 individuals were enrolled in the study (*Figure S1*). Of those enrolled, 34 participants were excluded for the following reasons: mild cognitive impairment (*n* = 8), acute pain (*n* = 5), lack of time to finish the tests (*n* = 5), voluntary abandonment (*n* = 3), depression (*n* = 5), unstable clinical condition (*n* = 3), Covid-19 infection sequalae (*n* = 2), neurological sequelae (*n* = 2) and insufficient visual acuity (*n* = 1); 8 others were excluded because of inconsistent explosive contractions (5 at rest and 3 during the fatigability period).

Thus, 117 participants were considered for analysis (*Table 1*). The a priori pairing of individuals in terms of age and prevalence of diabetes was verified, with no significant difference between CKD patients and controls. The biochemical data regarding CKD individuals are shown in *Table S1*.

Table 1 Baseline data in the study population

|  | All | Group | | P-values |
|---|---|---|---|---|
|  |  | CKD | Controls |  |
| n | 117 | 57 | 60 |  |
| **Anthropometric data** |  |  |  |  |
| Age (years), median (IQR) | 69 (9) | 69 (11) | 69 (11) | 0.156 |
| Gender (% females) | 37.6% | 29.8% | 45.0% | 0.090 |
| BMI (kg/m$^2$), median (IQR) | 27.6 (6.2) | 27.9 (5.4) | 26.8 (5.9) | 0.106 |
| Charlson, median (IQR) | 5 (4) | 7 (2) | 3 (2) | **<0.001** |
| SGA, n (%) |  |  |  | 0.999 |
| A | 113 (96.6%) | 55 (96.5%) | 58 (96.7%) |  |
| B | 4 (3.4%) | 2 (3.5%) | 2 (3.3%) |  |
| Diabetes, n (%) | 49 (41.9%) | 24 (42.1%) | 25 (41.7%) | 0.962 |
| Blood pressure (mmHg), median (IQR) |  |  |  |  |
| Systole | 140 (23) | 143 (23) | 130 (24) | **0.036** |
| Diastole | 74 (11) | 77 (10) | 70 (13) | 0.238 |
| Heart rate (b.p.m.), median (IQR) | 69 (16) | 69 (15) | 70 (15) | 0.406 |
| Dominant arm (% right-handed) | 98.3% | 96.5% | 100% | 0.235 |
| **Biochemical data**[a] |  |  |  |  |
| Haemoglobin (mg/dL), median (IQR) | 13.2 (3.0) | 12.3 (2.1) | 14.6 (1.8) | **<0.001** |
| Sodium (mmol/L), median (IQR) | 141 (3) | 140 (3) | 141 (3) | 0.665 |
| Potassium (mmol/L), median (IQR) | 4.4 (0.6) | 4.5 (0.6) | 4.4 (0.6) | 0.874 |
| Creatinine (μmol/L), median (IQR) | — | 198 (154) | 67 (19) | **<0.001** |
| GFR CKD-EPI (mL/min/1.73 m$^2$), median (IQR) | — | 25 (20) | 88 (12) | **<0.001** |

Bold P-values are considered statistically significant.
Abbreviations: BMI, body mass index; CKD, chronic kidney disease; GFR CKD-EPI, glomerular filtration rate with the Chronic Kidney Disease Epidemiology Collaboration equation; IQR, interquartile range; SGA, subjective global assessment.
[a]Only the biochemical data with more than 75% of available data are shown.

## Propensity-score-matched groups

A total of 100 individuals (66 males and 34 females) were finally considered in the propensity-score-matched groups (*Table 2*). The group for each gender was well balanced, except for haemoglobin and Charlson Comorbidity Index. In addition, MVF was lower in CKD patients than controls both in males ($P = 0.024$, Cohen's $d = 0.570$) and in females ($P = 0.002$, Cohen's $d = 1.190$).

## Zero-dimensional analysis

### t-RFD$_{peak}$

Fatigability delayed the t-RFD$_{peak}$ ($P = 0.007$; $\eta_p^2 = 0.032$), with a significant effect observed at the second and third explosive contractions performed during the fatigability period (*Figure 3A*), without time interaction with groups ($P = 0.097$) or gender ($P = 0.726$). The representation between genders is shown in *Figure S2*.

### RFD$_{peak}$

The RFD$_{peak}$ was significantly reduced at each time point ($P < 0.001$; $\eta_p^2 = 0.165$; *Figure 3B*) and showed interaction with gender (for $P < 0.001$; $\eta_p^2 = 0.053$), but the decrement during time failed to show significant interaction with group (for $P = 0.077$). RFD$_{peak}$ was lower in females than in males and also in CKD subjects than in controls ($P < 0.001$; $\eta_p^2 = 0.273$ and $P = 0.005$; $\eta_p^2 = 0.079$, respectively).

### NRFD$_{peak}$

The NRFD$_{peak}$ decreased significantly with time ($P < 0.001$; $\eta_p^2 = 0.165$; *Figure 3C*) and an interaction with group was noted (for $P = 0.009$; $\eta_p^2 = 0.034$) with a significant greater decrease in NRFD$_{peak}$ in CKD patients, but this was unrelated to gender ($P = 0.056$). For males, the relative decrement in NRFD$_{peak}$ shown with the fatigability protocol was 22.6% in controls and 37.6% in CKD patients, while for females, there was a 13.1% reduction in controls and a 26.9% reduction in CKD patients.

### MVF-RFD$_{peak}$

The MVF-RFD$_{peak}$ decreased significantly with time ($P < 0.001$; $\eta_p^2 = 0.046$) from the third explosive contraction performed during the fatigability period to the last (*Figure 3D*), and females had higher MVF-RFD$_{peak}$ than males ($P = 0.039$; $\eta_p^2 = 0.044$).

### Time period RFD

Each absolute and normalized RFD considered at fixed time points significantly decreased with the fatigability protocol (for both $P < 0.001$; $\eta_p^2 = 0.232$ and 0.235, respectively). Contrary to the NRFDs, RFDs show sex interaction with fatigability ($P < 0.001$; $\eta_p^2 = 0.045$).

In absolute RFD, CKD patients and controls showed different time–group interactions according to the time point considered ($P = 0.003$; $\eta_p^2 = 0.063$). In the early phase of contraction (≤100 ms from onset), CKD patients had lower RFD$_{75}$ and RFD$_{100}$ than controls (for both $P < 0.001$; $\eta_p^2 = 0.068$ and

Table 2  Baseline data for propensity-score-matched groups by gender

| | Gender | | | | | | | | |
|---|---|---|---|---|---|---|---|---|---|
| | Male | | | | Female | | | | |
| | All | CKD | Controls | P-values | All | CKD | Controls | P-values | |
| n | 66 | 33 | 33 | | 34 | 17 | 17 | | |
| **Anthropometric data** | | | | | | | | | |
| Age (years), median (IQR) | 71 (10) | 71 (10) | 70 (9) | 0.096 | 67 (6) | 67 (5) | 66 (7) | 0.468 | |
| BMI (kg/m²), median (IQR) | 27.8 (4.6) | 27.4 (4.2) | 28.0 (4.6) | 0.994 | 27.2 (7.1) | 28.9 (9.6) | 25.6 (7.2) | 0.274 | |
| Charlson, median (IQR) | 5 (3) | 7 (2) | 4 (2) | <**0.001** | 4 (3) | 6 (2) | 3 (2) | <**0.001** | |
| SGA, n (%) | | | | | | | | 0.999 | |
| A | 100% | 100% | 100% | — | 33 (97.1%) | 16 (94.1%) | 17 (100%) | | |
| B | 0 | 0 | 0 | | 1 (2.9%) | 1 (5.9%) | 0 | | |
| Diabetes, n (%) | 31 (47.0%) | 13 (39.4%) | 18 (54.6%) | 0.218 | 13 (38.2%) | 6 (35.3%) | 7 (41.2%) | 0.724 | |
| Blood pressure (mmHg), median (IQR) | | | | | | | | | |
| Systole | 140 (20) | 140 (20) | 130 (22) | 0.052 | 140 (25) | 145 (16) | 130 (23) | 0.306 | |
| Diastole | 76 (11) | 78 (10) | 70 (15) | 0.072 | 70 (10) | 70 (8) | 70 (11) | 0.999 | |
| Heart rate (b.p.m.), median (IQR) | 70 (15) | 69 (14) | 72 (15) | 0.131 | 70 (12) | 70 (14) | 70 (11) | 0.865 | |
| Dominant arm (% right-handed) | 98.5% | 97.0% | 100% | 0.999 | 97.1% | 94.1% | 100% | 0.999 | |
| MVF (N), median (IQR) | 293.5 (92.8) | 270.0 (77.7) | 307.2 (83.4) | **0.024** | 199.5 (60.3) | 174.5 (48.5) | 212.1 (37.3) | **0.002** | |
| **Biochemical data** | | | | | | | | | |
| Haemoglobin (mg/dL), median (IQR) | 13.8 (3.1) | 12.4 (2.0) | 15.1 (1.4) | <**0.001** | 12.7 (2.0) | 11.7 (1.3) | 13.5 (0.9) | **0.027** | |
| Sodium (mmol/L), median (IQR) | 140 (3) | 141 (2) | 140 (4) | 0.564 | 141 (5) | 141 (6) | 141 (4) | 0.656 | |
| Potassium (mmol/L), median (IQR) | 4.5 (0.7) | 4.5 (0.6) | 4.3 (0.7) | 0.485 | 4.4 (0.5) | 4.3 (0.4) | 4.6 (0.5) | 0.142 | |
| Creatinine (μmol/L), median (IQR) | 114 (151) | 233 (140) | 77 (14) | <**0.001** | 89 (84) | 153 (80) | 62 (14) | <**0.001** | |
| eGFR CKD-EPI (mL/min/1.73 m²), median (IQR) | 55 (63) | 24 (15) | 88 (12) | <**0.001** | 55 (53) | 34 (21) | 88 (9.5) | <**0.001** | |

Bold P-values are considered statistically significant.
Abbreviations: BMI, body mass index; CKD, chronic kidney disease; eGFR CKD-EPI, estimated glomerular filtration rate with the Chronic Kidney Disease Epidemiology Collaboration equation; IQR, interquartile range; MVF, maximum voluntary force; N, Newton; SGA, subjective global assessment.

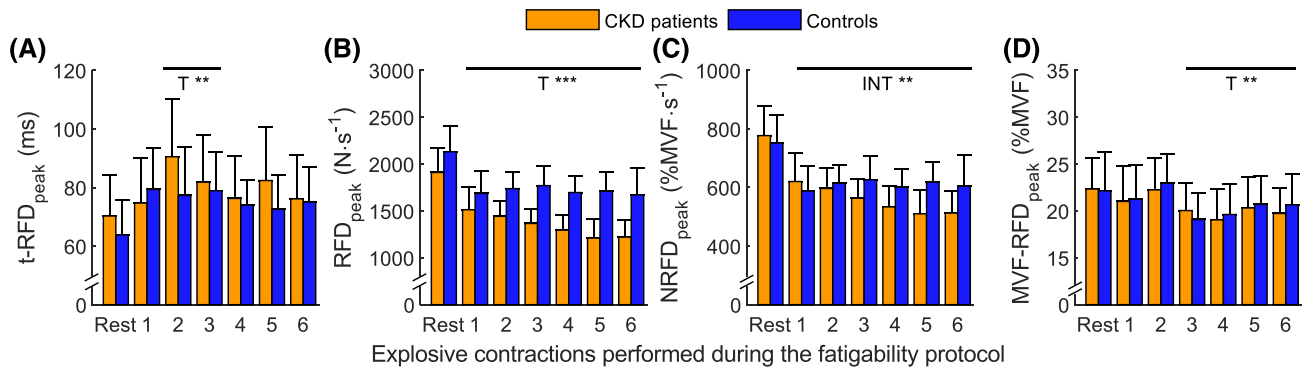

Figure 3 Scalars evolution during the fatigability protocol (mean ± 95% confidence interval) in the two groups. (A) t-RFD$_{peak}$: time-to-peak of the rate of force development (RFD) relative to onset of contraction; (B) RFD$_{peak}$: peak of the RFD–time curve; (C) NRFD$_{peak}$: maximal voluntary force (MVF)-normalized RFD$_{peak}$; (D) MVF-RFD$_{peak}$: percentage of MVF at RFD$_{peak}$. Horizontal black lines on the graphs represent within effect of the post hoc of the ANOVA, for TIME (T) or INTERACTION between group and time (INT). See Figure S2 for the representation of each gender. * for $P < 0.05$, ** for $P < 0.01$, *** for $P < 0.001$.

0.064, respectively). In the later phase of contraction, CKD patients remained lower than controls regarding the RFD$_{150}$ and RFD$_{200}$ ($P = 0.001$ and $P = 0.019$, respectively; for both $\eta_p^2 = 0.060$). The RFD at fixed time periods is lower in females than males ($P < 0.001$; $\eta_p^2 = 0.324$) and lower in CKD patients than controls ($P < 0.001$; $\eta_p^2 = 0.137$). When normalized to MVF, no differences were found, neither between genders ($P = 0.433$) nor between groups ($P = 0.067$). No relationships were found between scalars and eGFR.

### One-dimensional analysis

In males, as for all the one-dimensional absolute force estimates (Figure 4A), the MVF-normalized force–time curves show an interaction between group and the fatigability protocol in the early and later phases of contraction (46–82 and 127–200 ms; Table S2 and Figure 4B). The post hoc analysis revealed early neuromuscular fatigability in controls compared with CKD patients. All estimates, both absolute and MVF normalized, were significantly decreased in controls starting from the first explosive contraction in the fatigability protocol. For CKD patients, there was a delayed reduction in these force estimates compared with controls, evident at the fourth explosive contraction regarding the force–time curves (Figure S3a) and at the second explosive contraction regarding impulse, RFD, MVF-normalized force–time curves, MVF-normalized impulse and MVF-normalized RFD (Figure S3b,c).

Controls manifested an RFD impairment in the increasing phase and, in the reduction phase, in the rate in rise of contractile force, with an insufficient decrement in muscular performances with time to involve the RFD$_{peak}$ period. However, and albeit delayed, CKD patients had a reduction the MVF-normalized RFD–time curves in the RFD$_{peak}$ period lasting 42–85 ms to the 34–139 ms time window at the end of the fatiguing task.

In females, CKD females had lower absolute force and impulse (Figure 4C) than female controls, but no difference was noticed when normalized (Figure 4D). No differences emerged in the post hoc analysis (Figure S4a–c), except for the MVF-normalized RFD–time curves, in which a reduction with time was found in controls in the 133–161 ms time window, while there was no decrease in muscle performances in CKD patients.

### Surface electromyography analysis

#### At rest

Two outliers were removed from analysis (one male CKD patient, 67 years old with 5.7 of $z$-score, and one male control, 79 years old with 3.5 of $z$-score). At rest, the integrated sEMG was lower in CKD patients than controls (Figure 5) in the 0–100 ms (4.3 ± 2.4 vs. 5.5 ± 2.7%EMG$_{max}$·s, respectively; $P = 0.049$, Cohen's $d = 0.458$) and 0–200 ms (10.3 ± 4.7 vs. 12.6 ± 4.7%EMG$_{max}$·s, respectively; $P = 0.016$, Cohen's $d = 0.496$) time windows, but not for the others (0–30 ms: $P = 0.160$; 0–50 ms: $P = 0.087$). In males, only the 0–200 ms time window was significantly lower in CKD patients than controls (10.9 ± 5.3 vs. 13.9 ± 4.6%EMG$_{max}$·s, respectively; $P = 0.018$, Cohen's $d = 0.606$). No differences were noticed in females.

#### Fatigability effect

All the sEMG parameters decreased with fatigability ($P < 0.001$; $\eta_p^2 = 0.350$). No interaction between group–fatigability or sex–fatigability was shown regarding the evolution of sEMG parameters during the fatigability exercise, independently of the time window considered ($P = 0.257$; $\eta_p^2 = 0.014$ and $P = 0.122$; $\eta_p^2 = 0.020$, respectively). Overall evolution of sEMG time windows (from 0–30 to 0–200 ms, independently of the fatigability process) seems to show group–time window ($P = 0.055$; $\eta_p^2 = 0.038$) or sex–time window ($P = 0.076$;

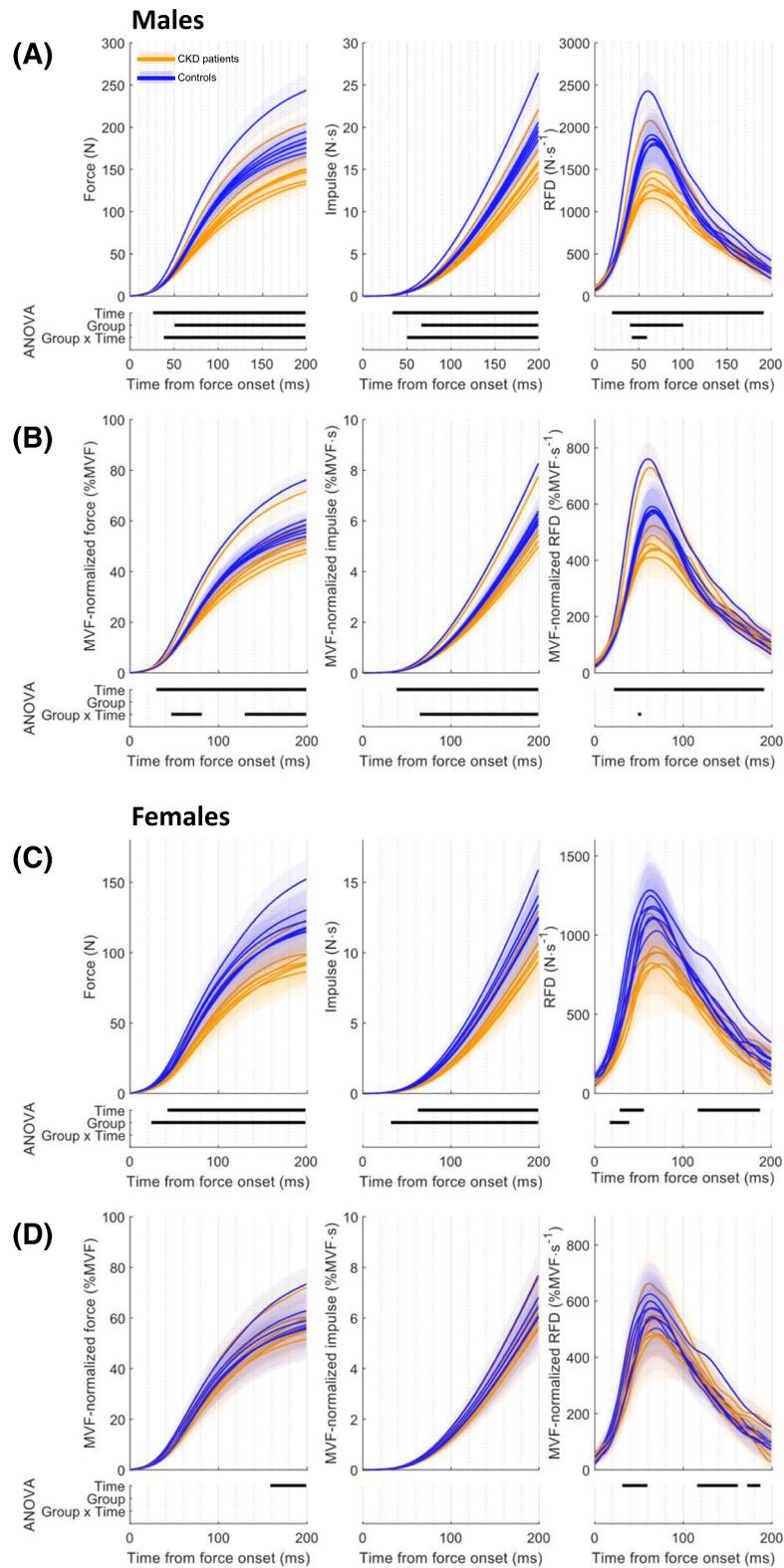

**Figure 4** Explosive force estimates comparison of chronic kidney disease (CKD) patients and controls in males (absolute data (A) and maximum voluntary force [MVF] normalized (B)) and in females (absolute data (C) and MVF normalized (D)). The shaded area represents 95% confidence intervals. The results of the ANOVA by means of statistical parametric mapping are displayed in the lower panel. The statistically significant time windows are represented by the horizontal black line for Time, Group and Group × Time interaction, in the first, second and third rows, respectively. RFD, rate of force development.

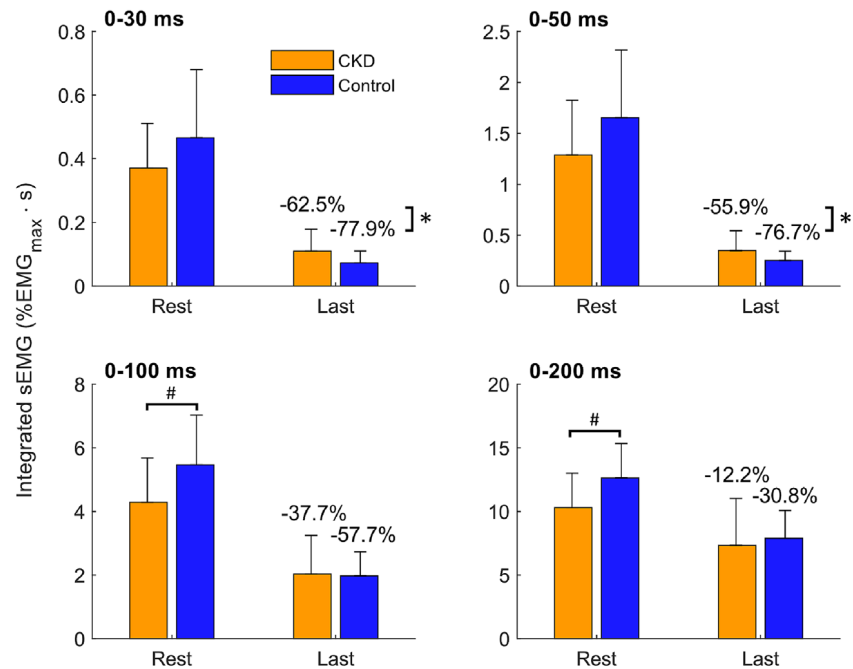

**Figure 5** Evolution of the integrated surface electromyography (sEMG) (mean ± 95% confidence interval) between chronic kidney disease (CKD) patients and controls at different time intervals. *Between-group difference of integrated sEMG decrease compared with rest, for $P < 0.05$. #Between-group difference of integrated sEMG at rest, for $P < 0.05$.

$\eta_p^2 = 0.033$) interaction after the Greenhouse–Geisser correction. Post hoc analysis identifies statistically significant lower muscle activity within the 0–200 ms time window in CKD patients compared with controls and in females compared with males ($P = 0.006$, Cohen's $d = 0.421$; $P = 0.018$, Cohen's $d = 0.376$, respectively).

Fatigability–time window interaction was evidenced ($P < 0.001$; $\eta_p^2 = 0.185$). Thus, a pre- (i.e., at rest) or post- (i.e., at the end of exercise) comparison was performed. Controls showed greater decrease of sEMG than CKD patients in the 0–30 ms ($-77.9 \pm 21.3\%$ vs. $-62.5 \pm 34.6\%$, respectively; $P = 0.020$, Cohen's $d = 0.533$) and 0–50 ms ($-76.7 \pm 22.6\%$ vs. $-55.9 \pm 39.9\%$, respectively; $P = 0.010$, Cohen's $d = 0.640$) time windows (0–100 ms: $P = 0.251$; 0–200 ms: $P = 0.794$). Control males had greater reduction of sEMG than CKD males in the 0–50 ms time window ($-78.0 \pm 25.4\%$ vs. $-55.4 \pm 43.9\%$, respectively; $P = 0.035$, Cohen's $d = 0.631$), and no differences were found in females.

## Discussion

The main results are that CKD patients had higher neuromuscular fatigability than controls, with a delayed onset of neuromuscular fatigability and an impairment in the early phase of contraction associated with an impairment in the motor-unit recruitment. The present study used explosive force estimates to assess neuromuscular fatigability in elderly CKD patients coupled with sEMG analysis. To the best of our knowledge, this is the first study that shows that pre-dialysis CKD patients have increased neuromuscular fatigability compared with an age- and diabetes-matched control group. In addition, SPM analysis evidenced delayed task-induced neuromuscular fatigability in CKD patients. These impairments may be associated with a disturbance in motor-unit recruitment during explosive contractions.

The MVF-$RFD_{peak}$ in our study (22.1% MVF in males and 24.2% MVF in females) is in accordance with Rodríguez-Rosell et al. (2018), who reported approximately 30% MVF in their review. Moreover, our study found a mean delay of 67 ms from onset of contraction to $RFD_{peak}$ at rest, which is in accordance with studies involving a wide range of ages and tasks, showing a mean delay from 52 to 74 ms.[27,32,33] Thus, the force signal assessed and the parameters extracted in elderly CKD patients are in accordance with the state of the art.

The MVF normalization is justified because the lower $RFD_{peak}$ in CKD patients is mainly influenced by the difference in MVF capacities. Thus, the $NRFD_{peak}$ has a different interaction with time: Regardless of their MVF, the peak in the rate in the rise of contractile force in CKD patients is more affected by the individualized submaximal fatiguing task than it is in controls. In controls, $NRFD_{peak}$ decreases over time by 22.6% in males and 13.1% in females, whereas CKD patients manifest a 37.6% and a 26.9% reduction in males and females, respectively. This statistically and clinically relevant difference is evidence of higher neuromuscular fatigability in CKD patients than in controls.

SPM analysis supports the results obtained from the scalars, showing higher neuromuscular fatigability in CKD patients regarding the peak of the RFD–time curve (in absolute and MVF normalized). Controls did not manifest a decrement in RFD–time curves in the $RFD_{peak}$ period whereas CKD patients did when compared with rest. Thus, both CKD patients and controls manifested neuromuscular fatigability, while the amount of decrement was higher in CKD patients at $RFD_{peak}$.

This result is not in accordance with a previous study showing that, compared with controls, elderly pre-dialysis CKD patients did not manifest higher neuromuscular fatigability during a 60% MVF submaximal fatigability protocol.[12] This can be explained by the submaximal threshold used, because the origins (i.e., central or peripheral) of neuromuscular fatigability are known to be task specific.[29,34] The difference may be due to the different muscles tested or may depend on the fact that RFD is a more sensitive means for assessing neuromuscular fatigability than time-to-task failure.[17] Different alterations in neuromuscular components can shed light on the higher neuromuscular fatigability in CKD patients noticed in this study.[10]

The handgrip test principally involves the *Flexor Digitorum Profundus*, *Superficialis* and, to a lesser degree, *Flexor Pollicis Longus*. These muscles are probably exposed to compliance changes explained by the reduction in myogenic differentiation associated with decreased eGFR, leading to muscular fibrosis and increased stiffness. However, this finding was not tested in CKD patients and the relationship with RFD performances remains uncertain. No relationship between muscle stiffness and RFD in older people was found.[35]

Changes in tendon stiffness can induce RFD discrepancies, but no such change has been demonstrated in CKD patients, presumably because the rate of force transmission through the tendon is extremely fast, requiring a considerably large change in the tendon structure to significantly disrupt RFD.[16,25]

Human type-II muscle fibres are faster than type-I fibres. In CKD patients, advanced glycation end products might induce a transition towards fast-to-low-speed muscle fibre type.[36] However, the literature reports conflicting results or no differences with controls regarding muscle fibre-type proportion or area.[12,37]

Nevertheless, CKD patients had important histopathological abnormalities[12] with the accumulation of toxic substances, including advanced glycation end products, which induces intramuscular fatty infiltration, reduction in muscle mass and reduction in mitochondrial efficiency.[36]

As blood flow is similar at rest and during a sustained isometric handgrip task at 40% MVF,[38] and due to the intermittent design of the fatigability protocol, we conclude that oxygen supply is not the major limitation in CKD patients.

According to the literature, the composition and size of muscle fibre, musculotendinous structures and oxygen supply are not deeply affected in CKD patients. Thus, the factor mainly responsible for the higher neuromuscular fatigability is unlikely to be found at the peripheral level.

One-dimensional explosive force estimates (i.e., force, impulse and RFD–time curves) show that CKD patients manifested a delayed neuromuscular fatigability onset compared with controls. We can therefore speculate that CKD individuals have a diminished capacity to fully activate their muscles, leading to a discrepancy between the MVF performed and their real maximum evocable force capacities.[5] Lower MVF leads to performance by CKD individuals in the fatigability period that is below the 40% MVF threshold of their maximal capacities. This observation is a common result in the scope of chronic diseases.[7,39] This study highlights an impairment in explosive force generation in CKD patients compared with controls with exercise. Thus, it may be important to consider explosive force generation capacities during rehabilitation protocols in order to improve functional performances of these patients.

The early phase of RFD, linked to motoneuron discharge rate and speed of recruitment,[20] is impaired in CKD patients and this suggests an alteration in the efferent neural drive.[23] This hypothesis is supported by sEMG analysis in which CKD patients showed reduced capacities to recruit efficiently their motor units compared with controls. Temporal analysis of the time-integrated sEMG parameters failed to show group–time interaction, probably due to the high variability of these parameters. Nevertheless, the effect of CKD on the evolution of these parameters during exercise is considered as small, emphasizing their interest for future research. This result suggests that CKD induces an impairment of one or more processes of the descending motor command. In this context, the early neuromuscular fatigability associated with a greater decrease of the integrated sEMG noticed in controls could be explained by their higher capacities in muscle recruitment at rest compared with CKD patients.

It has been previously shown that CKD patients not on dialysis display an alteration of the nerve function.[40] This disease-related impairment is due to a decreased nerve conduction velocity secondary to an alteration of the axonal function.[40] Poor axonal function is multifactorial and its potential determinants include accumulation of advanced glycation end products,[36] hyperparathyroidism,[41] hyperkalaemia,[42] erythropoietin deficiency[43] or alteration of the $Na^+,K^+$-pump activity.[44] In addition, our results are similar to those obtained in other diseases such as cancer, where neuromuscular fatigability originates from cortical limitation during exercise, and patients are unable to recruit as much muscle fibres as controls, exacerbating neuromuscular fatigability.[7,8] We may hypothesize that both peripheral and central nervous systems are sensitive to mild and initial metabolic derangement during CKD, increasing the neuromuscular fatigability.[10]

This study included a relatively large sample permitting accurate pairing of individuals for age and diabetes prevalence. The differences across gender are partly related with a lower statistical power attributable to the unbalanced prevalence in the nephrology unit where patients were enrolled. Age-related changes in neuromuscular function differ in males and females, with different muscle contractile properties and a higher reduction in neural drive in males,[25] justifying separate analysis. The choice of a single assessment per patient was determined because multiple visits are expected to lead to a high drop-out rate in elderly volunteers, who often reside far from the hospital. Furthermore, the advantages of multiple visits are counterbalanced by a rapid deterioration in clinical condition. Unlike other studies, we focused on the assessment of muscle performance using handgrip, as this is considered reliable and implementable in clinical practice.[3] Above all, it reduces the biases resulting from lower limb vascular or neuropathic involvement in CKD patients.[40] Despite that twitch interpolation technique is considered as the gold standard in fatigability aetiology assessment during sustained maximal voluntary contractions, this method did not allow assessment of fatigability aetiology during rapid force capacities.[22] Moreover, our study was conceived to minimally interfere with patient well-being and this method is painful, leading to low benefit/risk ratio. The aforementioned reasons justified the use of non-invasive sEMG, more adapted to assess impairment of central activation in the initial phase of contraction. Finally, spontaneous physical activity was not recorded due to the time constrain of the experimentations in the routine care and because it did not significantly impact neuromuscular fatigability in the elderly.[45]

## Conclusions

Our study shows that CKD patients have higher neuromuscular fatigability and an impairment in motor-unit recruitment than controls matched by age, sex and diabetes. These results integrate the well-known reduction in muscle strength in CKD patients and open the way to new therapeutic and research pathways. Additional studies using artificial muscle stimulation testing in a larger CKD cohort are needed to determine the origin of these findings.

## Acknowledgements

The authors would like to thank all the volunteers who participated. We are grateful to the French Ministry of Training and Research for their partial financial support through Convention Industrielle de Formation par la Recherche (CIFRE) grant (No. 2018/1255). The authors also thank Susan Finnel for proofreading the text and Nicolas Peyrot and Sebastien Boyas for their advices. The authors of this manuscript certify that they comply with the ethical guidelines for authorship and publishing in the Journal of Cachexia, Sarcopenia and Muscle.[46]

## Conflict of interest statement

The authors declare that they have no conflicts of interest.

## Online supplementary material

Additional supporting information may be found online in the Supporting Information section at the end of the article.